%% file: paper.tex
\documentclass[sigconf, screen]{acmart}
\AtBeginDocument{%
  \providecommand\BibTeX{{%
    \normalfont B\kern-0.5em{\scshape i\kern-0.25em b}\kern-0.8em\TeX}}}

\acmPrice{}

\copyrightyear{2024}
\acmYear{2024}
\setcopyright{rightsretained}
\acmConference[ICSE '24]{2024 IEEE/ACM 46th International Conference on
Software Engineering}{April 14--20, 2024}{Lisbon, Portugal}
\acmBooktitle{2024 IEEE/ACM 46th International Conference on Software
Engineering (ICSE '24), April 14--20, 2024, Lisbon,
Portugal}\acmDOI{10.1145/3597503.3639116}
\acmISBN{979-8-4007-0217-4/24/04}

\usepackage{algorithmic}
\usepackage{graphicx}
\usepackage{textcomp}

\usepackage{xcolor}
\usepackage[T1]{fontenc} 
\usepackage{mathtools, nccmath}
\usepackage{algorithmic}
\usepackage{graphicx}
\usepackage{textcomp}
\usepackage{xcolor}
\def\BibTeX{{\rm B\kern-.05em{\sc i\kern-.025em b}\kern-.08em
    T\kern-.1667em\lower.7ex\hbox{E}\kern-.125emX}}
    
\usepackage{color,xcolor}
\usepackage{multirow}
\usepackage{colortbl}
\usepackage{booktabs} 
\usepackage{comment}
\usepackage{tablefootnote}
\usepackage{ifthen}
\usepackage{url}
\usepackage{longtable}
\usepackage{threeparttable}
\usepackage{hyphenat}
\usepackage{bbding}
\usepackage{xspace}
\usepackage[T1]{fontenc}
\usepackage[ruled,linesnumbered]{algorithm2e}
\usepackage{array,multirow,graphicx}
\usepackage{float}
\usepackage{balance}
\usepackage{hyperref}
\usepackage{tikz}
\usepackage{calc}
\usepackage{subfigure}
\usepackage{listings}
\usepackage{pifont}
\usepackage{url}
\usepackage{balance}
\usepackage{xspace}
\usepackage{hyperref,epsfig,endnotes}
\usepackage{array}
\usepackage{paralist}
\usepackage{multirow,makecell}
\usepackage[normalem]{ulem}
\usepackage{pgfplots}
\usepackage{enumitem}
\usepackage{tcolorbox}
\usepackage{courier}
\usepackage{enumitem}
\usepackage{soul}
\usepackage{comment}
\usepackage{url}

\usepackage{breakurl}

\newcommand{\tech}{\mbox{\textsc{VGX}}}
\def\BibTeX{{\rm B\kern-.05em{\sc i\kern-.025em b}\kern-.08em
    T\kern-.1667em\lower.7ex\hbox{E}\kern-.125emX}}

\newboolean{showcomments}
\setboolean{showcomments}{true}

\ifthenelse{\boolean{showcomments}}
{
\newcommand{\modify}{\textcolor{black}}

\newcommand{\revise}{\textcolor{black}}
\newcommand{\sw}[1]{}
\newcommand{\swsecond}[1]{}
\newcommand{\hr}[1]{}
\newcommand{\hrmark}[1]{}
\newcommand{\hrdel}[1]{}
\newcommand{\ques}[1]{}

\newcommand{\hrcrcdel}[1]{\sout{#1}}
}
{
\newcommand{\modify}{\textcolor{black}}

\newcommand{\revise}{\textcolor{black}}
\newcommand{\sw}[1]{}
\newcommand{\swsecond}[1]{}
\newcommand{\hr}[1]{\textcolor{green}{#1}}
\newcommand{\hrmark}[1]{}
\newcommand{\hrdel}[1]{}
\newcommand{\ques}[1]{}

\newcommand{\hrcrcdel}[1]{}
} 

\emergencystretch=\maxdimen

\begin{document}

\title{{\tech}: Large-Scale Sample Generation for Boosting Learning-Based Software Vulnerability Analyses}

\author{Yu Nong}
\orcid{0000-0002-8598-5181}
\affiliation{
 \institution{Washington State University}
 \country{}
 }
\email{yu.nong@wsu.edu}

\author{Richard Fang}
\orcid{0009-0009-8104-9917}
\affiliation{
 \institution{Washington State University}
 \country{}
 }
\email{richardfang2005@gmail.com}

\author{Guangbei Yi}
\orcid{0009-0009-6679-5153}
\affiliation{
 \institution{Washington State University}
 \country{}
 }
\email{guangbei.yi@wsu.edu}

\author{Kunsong Zhao}
\orcid{0000-0001-9886-0460}
\affiliation{%
 \institution{Hong Kong Polytechnic University}
 \country{}}
\email{kunsong.zhao@connect.polyu.hk}

\author{Xiapu Luo}
\orcid{0000-0002-9082-3208}
\affiliation{%
 \institution{Hong Kong Polytechnic University}
 \country{}}
\email{csxluo@comp.polyu.edu.hk}

\author{Feng Chen}
\orcid{0000-0002-4508-5963}
\affiliation{%
 \institution{The University of Texas at Dallas}
 \country{}}
\email{feng.chen@utdallas.edu}

\author{Haipeng Cai}
\orcid{0000-0002-5224-9970}
\authornote{Haipeng Cai is the corresponding author.}
\affiliation{
 \institution{Washington State University}
 \country{}
 }
\email{haipeng.cai@wsu.edu}


\input{abstract-cai}



\maketitle

\input{intro-cai}
\input{motivation-cai}

\input{approach-compressed}

\input{evaluation}

\input{discussion}

\input{related}
\input{conclusion}

\section*{Acknowledgment}
We thank the reviewers for their constructive comments, 
which helped us improve our paper. 
This research was supported by the Army Research Office (ARO) through grant W911NF-21-1-0027. 

\bibliographystyle{ACM-Reference-Format}
\bibliography{paper}


\end{document}

%% file: abstract-cai.tex

\begin{abstract}
Accompanying the successes of learning-based defensive 
software vulnerability analyses 
is the lack of large and quality sets of labeled vulnerable program samples, which 
impedes further advancement of those defenses. 
Existing automated sample generation approaches have shown potentials yet still fall short of practical expectations due to the 
high noise in the generated samples. 
This paper proposes {\tech}, a new 
technique aimed for \textit{large-scale} generation of \textit{high-quality} vulnerability datasets.  
Given a normal program, {\tech} identifies the code contexts in which 
vulnerabilities can be injected, using a customized Transformer featured with 
a new value-flow-based position encoding and pre-trained 
against 
new 
objectives 
particularly for learning 
code structure and context. 
Then, {\tech} materializes vulnerability-injection code editing in the identified contexts using patterns of such edits obtained from both 
historical fixes and human knowledge about real-world vulnerabilities. 

Compared to four state-of-the-art (SOTA) (i.e., pattern-, Transformer-, GNN-, and pattern+Transformer-based) baselines, 
{\tech} achieved 99.09-890.06\% higher F1 
and \modify{22.45\%-328.47\%} higher label accuracy. 
For in-the-wild sample production, {\tech} generated 150,392 vulnerable samples, from which we randomly chose 10\% to assess how much 
these samples help vulnerability detection, localization, and repair. 
Our results show SOTA techniques for these three application tasks achieved 
19.15--330.80\% higher F1, 12.86--19.31\% higher top-10 accuracy, and 85.02--99.30\% higher top-50 accuracy, respectively, by adding those samples to their original training data. 
These samples also helped a SOTA vulnerability detector discover 13 more real-world vulnerabilities (CVEs) in critical systems (e.g., Linux kernel) that would be 
missed by the original model. 

\end{abstract}


%% file: intro-cai.tex
\vspace{0pt}
\section{Introduction}\label{sec:intro}
Modern software  
is widely afflicted by security vulnerabilities~\cite{lipolyfuzz,li2022polycruise,fu2021flowdist,fu2019dynamic}, 
which are increasingly consequential~\cite{vuleffects2013,li2023pyrtfuzz}. 
Thus, it is crucial to develop effective defenses against 
vulnerabilities, for which data-driven, especially deep-learning-based methods have demonstrated tremendous potential, including vulnerability detection~\cite{zhou2019devign,wu2022vulcnn,zou2022mvulpreter,zhou2022vulnerability}, localization~\cite{fu2022linevul,hin2022linevd,li2020vuldeelocator}, and repair~\cite{harer2018learning,fu2022vulrepair,chen2022neural}.

Yet accompanying the rising momentum of deep learning (DL) in software assurance 
is the glaring lack of quality training data. 
In fact, this problem has drawn growing attention 
from both academia and industry 
in recent years~\cite{chakraborty2022deep,zenong281412,industrydataquality1,industrydataquality2,industrydataquality3,nong2020preliminary,nong2021evaluating,cai2023generating}. 
Manual curation of such datasets is intuitively 
tedious hence clearly undesirable, and it can at best produce relatively small datasets hence unscalable~\cite{nvd,fan2020ac,wang2021patchdb,bhandari2021cvefixes}. 
Therefore, \textit{automatically} generating vulnerable samples at large scale and with high quality is of paramount significance. 

In response, a few data generation techniques have been developed, including neural code editing~\cite{yu2020study,dinella2020hoppity,yao2021learning,cai2023generating} and control-flow-based code stitching~\cite{mirskyvulchecker}. 
However, the former suffers a major chicken-egg problem---training the neural code editor requires a large set of quality vulnerable samples which are what we are lacking. The latter, by inserting artificial vulnerable code patterns to real-world code, results in unrealistic 
vulnerabilities which only have
limited usage. 
Other, even earlier approaches~\cite{zheng2021d2a,wen22fse,wen22fsetool,zenong281412}, including adaptable ones originally designed for program bug repair~\cite{bader2019getafix,dinella2020hoppity} 
are subject to even greater limitations (e.g., higher noise~\cite{yu2020study} and lower coverage of vulnerability classes~\cite{zhang2021framework}). 
Lately, VulGen~\cite{nongvulgen} made good progress; yet it still suffers from high noise in the generated data due to its low generation accuracy, as well as overfitting to the seed vulnerable samples it learns from.
In this paper, aiming at \textit{large-scale} generation of \textit{high-quality} vulnerable program samples, we developed an advanced vulnerability injection technique, named {\tech} (short for \textbf{V}ulnerability \textbf{G}eneration e\textbf{X}panded). 
{\tech} combines \modify{Step} (1) \emph{semantics-aware contextualization}, which identifies the context of vulnerability-injection code edits, with \modify{Step} (2) \emph{human-knowledge-enhanced edit pattern formation}, using the materialized contextualized code edits. 

In Step \textbf{(1)}, {\tech} builds a Transformer-based contextualization model by designing a novel attention mechanism with absolute and relative position encoding both based on value flow relationships among code tokens, and then pre-training the customized Transformer on a large code corpus, followed by fine-tuning it on \textit{an existing (task-specific) 
dataset} of vulnerability-introducing code locations. 
To benefit more from the pre-trained model, 
{\tech} also introduces new pre-training objectives explicitly geared to our particular task (of contextualizing injection edits). 

In Step \textbf{(2)}, {\tech} starts with extracting vulnerability-introducing code edit patterns from the same task-specific dataset and then enriches the extracted patterns with manually identified, different patterns, followed by diversifying the enriched patterns through manually derived pattern mutation rules. Both the manual pattern and mutation rule definitions are based on human knowledge about existing real-world vulnerabilities garnered 
from CWE/CVEs on NVD~\cite{nvd} and bug/issue reports on GitHub. 

The semantics-informed design in Step (1) helps {\tech} achieve effective contextualization of vulnerability-injecting code edits, while the human knowledge incorporation in Step (2) helps {\tech} overcome potential overfitting to the small task-specific dataset.
In addition, we also (1) pre-train the programming language model on the ASTs of the input code corpus in order to better learn syntactic and structural information in programs and (2) improve the training with data augmentation (through semantics-preserving code refactoring), as inspired by earlier 
works~\cite{niu2022spt,mirskyvulchecker}.

\vspace{-1pt}
To assess {\tech}, we pre-trained its contextualization model on 1,214K C functions and fine-tuned it on 7K real-world vulnerability-fixing
samples augmented by 156K of their refactored versions. From these 7K samples along with manual pattern refinement and diversification, we obtained 604 vulnerability-introducing edit patterns. We then conducted five sets of experiments.

\vspace{-1pt}
In the \textbf{first} set, we evaluated the accuracy of {\tech} versus four state-of-the-art baselines (VulGen~\cite{nongvulgen}, CodeT5~\cite{wang2021codet5}, Getafix~\cite{bader2019getafix}, and Graph2Edit~\cite{yao2021learning}) against 775 testing (normal) samples with ground-truth (vulnerable versions). 
{\tech} achieves 59.46\% precision, 22.71\% recall, and 32.87\% F1 (239.77\%-1173.23\%, 44.28\%-780.23\%, 99.09\%-890.06\% higher than the baselines), when considering as true positives the generated vulnerable samples exactly matching ground truths. 
When counting all of the generated samples that are indeed vulnerable (i.e., not exactly matching the ground truth \modify{but still exploitable}) as success cases, 
{\tech} achieved a \modify{93.02\% success rate, 22.45\%-328.47\%} higher than the baselines. 
In the \textbf{second} set of experiments, we showed that each of the novel design elements of {\tech} (especially the value-flow-based semantics-aware attention mechanism and human-knowledge-based enhancement of edit patterns)
played a significant role in its overall performance merits. 

In the \textbf{third} experiment set, we deployed {\tech} in the wild on 738K samples, hence producing 150K vulnerable samples in 50 hours 48 minutes
with a \modify{90.13\% success rate, 118.07\% more accurate than VulGen---the best baseline}. 
Then, in the \textbf{fourth} set, we augmented the training sets of SOTA defensive vulnerability analyses with \modify{10\% of those 150K samples, considering the scalability of the analysis tools}. 
In this way, {\tech} helped improve (1) (function-level) vulnerability detection by 19.15\%-330.80\% in terms of F1; (2) (line-level) vulnerability localization by 12.86\%-19.31\% in terms of top-10 accuracy; and (3) vulnerability repair by 85.02\%-99.30\% in terms of top-50 accuracy. 
\modify{This is 4.19\%-266.42\% higher than adding VulGen's generated and existing synthetic samples.}
Finally, in the \textbf{fifth} set of experiments, we applied the augmented version of a SOTA vulnerability detector to 71 latest CVEs in real-world projects and found 62, including 13 that would be missed by the original model. 

\revise{While we currently implement hence evaluate {\tech} for C programs, our {\tech} approach/methodology is not limited to a particular programming language. The proposed contextualization and pattern mining techniques can be readily adapted for a different language, if given the AST and value flow parser for that language. The step of enhancing vulnerability-injection edit patterns based on human knowledge also follows a principled procedure, which can be reasonably replicated for other languages as well.}

In summary, this paper contributes the following:
\vspace{-5pt}
\begin{itemize}[leftmargin=11pt]
    \item A novel design for injection-based vulnerable sample generation that leverages \textit{code structure- and context-aware source code pre-training}, \textit{code semantics-aware Transformer attention}, data augmentation, and \textit{human knowledge} to overcome key limitations of existing peer approaches. 

    \item An implementation and evaluation of the design that shows its \textit{significantly superior performance over four SOTA baselines}, including the latest, best-performing peer technique.

    \item A \textit{large and quality set of 150K vulnerable samples} produced using the implementation, which comes with corresponding normal code, ground-truth vulnerability and locations, and high label accuracy, hence ready for public use. 

    \item Empirical evidences that show the practical usefulness of the generated dataset in terms of the \textit{substantial improvement it brought to vulnerability detection, localization, and repair}, as well as its ability to enable the discovery of real-world vulnerabilities (CVEs) that original models miss. 
\end{itemize}

\vspace{0pt}\noindent
{\bf Open science.}
Source code of {\tech} along with all of the experimental and resulting datasets are 
\revise{\href{https://zenodo.org/records/10443177}{\underline{available here}}}. 

%% file: motivation-cai.tex
\section{Background and Motivation}\label{sec:bkgmotive}
\begin{figure}[tp]
\centering
	\vspace{-0pt}
	\includegraphics[width=0.95\linewidth]{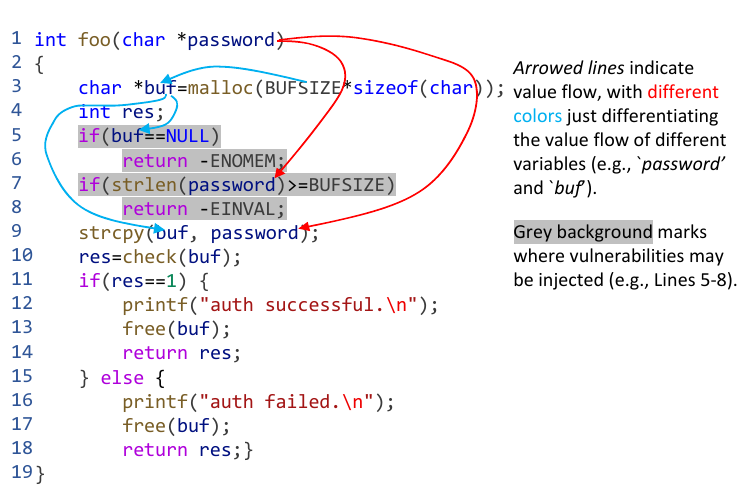}
    \vspace{-10pt}
	\caption{A motivating example on vulnerability injection.}
	\label{fig:illustrate}
	\vspace{-4pt}
\end{figure}

%
%
To automatically generate vulnerable samples, an intuitive idea is to (1) learn the patterns of known code edits that introduce vulnerabilities (i.e., \textit{vulnerability-introducing edits}) 
and then (2) apply to a normal program the patterns that are compatible with it, resulting in a vulnerable version of the program. Such an approach has been demonstrated to be meritorious in earlier works~\cite{bader2019getafix,yu2020study,nongvulgen}. 
Most recently, Nong et al. proposed VulGen~\cite{nongvulgen}, an injection-based vulnerable code generation technique that also leverages this strategy. Besides the learned patterns, it utilizes a Transformer-based localization model to 
locate 
\emph{where} to inject vulnerabilities. While an important step, VulGen still falls short of large-scale high-quality vulnerability data generation due to the following limitations:


\modify{\textbf{Limitation \textcircled{1}: The learned vulnerability-injection code edit patterns are limited to those seen in the small vulnerability dataset.} 
VulGen mines such patterns from existing vulnerability-introducing code edits. Yet since available datasets of such edits are (even collectively) limited, the patterns mined are a small subset of the extant knowledge about how real-world vulnerabilities can be introduced to software (as embodied in extant vulnerability documentation such as CWE/CVEs in NVD and issues/bug reports on GitHub).} 
For example, in Figure~\ref{fig:illustrate}, the statements marked gray at Lines 5-6 and 7-8 can be injected 
"null-pointer-dereference" 
and "buffer-overflow" vulnerabilities, respectively, by deleting the {\tt if}-statements. However, patterns as 
such may be \textit{too general} such that the condition and the return values in Lines 5 and 6 are not specified, or \textit{too specific} such that all the tokens for the condition in Line 7 are specified. 
Only the tokens "NULL", "ENOMEM", "EINVAL" are crucial for vulnerability injection in this example, but the respective patterns are difficult to mine 
due to the lack of available code samples covering them.
Thus, \textit{to leverage their full potential, the patterns need to better cover the extant knowledge, as can be achieved by incorporating human knowledge into the pattern mining process}. 

\begin{figure*}[tp]
\centering
\vspace{-0pt}
\includegraphics[width=0.95\linewidth]{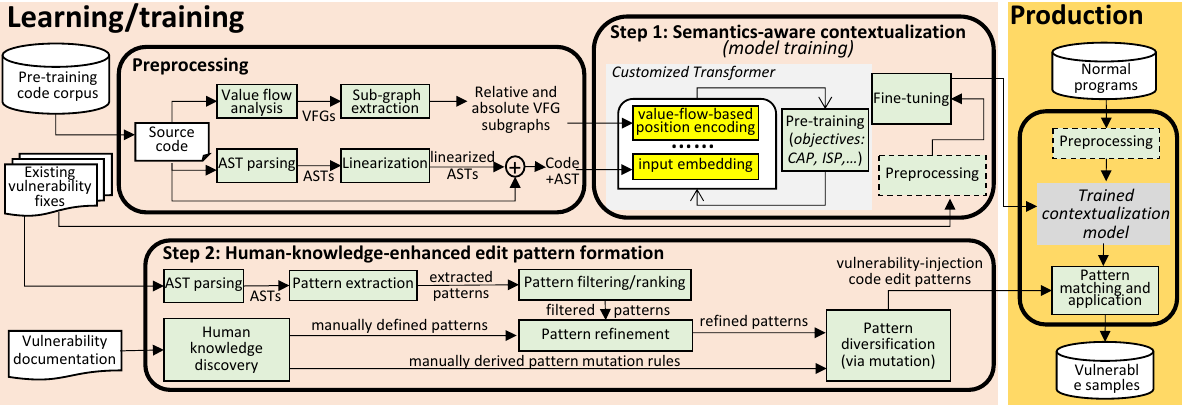}
\vspace{-10pt}
\caption{Design overview of the {\tech} approach, highlighting its two main phases: \textit{learning/training} and \textit{production}.}
\label{fig:overview}
\vspace{-4pt}
\end{figure*}

\modify{\textbf{Limitation \textcircled{2}:  The localization model is trained on source code as natural language token sequences, ignoring semantic information that is essential for accurately identifying where vulnerabilities may be injected in a given program. }
While the Transformer-based localization model in VulGen has shown to be reasonably capable of (e.g., faulty) code localization thanks to its explicit learning about locations through position encoding, the goal of effectively contextualizing vulnerability-injection code edits is hard to achieve with a vanilla Transformer position encoder. The reason is that code vulnerabilities are context-sensitive:} the same lines of code (e.g., copying an array to a buffer) may cause a vulnerability when placed in one code context (e.g., where there is no boundary check against the buffer's size) but would not lead to any vulnerable behavior in another context (e.g., where the buffer has been ensured to be large enough to hold the array). 
For example,
in Figure 1, the sample cannot be injected a "buffer-overflow" vulnerability by just removing Lines 7-8 if {\tt password[BUFSIZE-1]} has been set
to ’\textbackslash0’.
Thus, whether pattern-based injection is fruitful depends on whether the pattern is applied in the right code context. 
\modify{\textit{Identifying the right injection contexts is clearly reliant on code semantics, which can be learned by making the position encoding explicitly aware of semantic code information like value flows}~\cite{guo2020graphcodebert} (e.g., value flow of {\tt password} as illustrated in Figure~\ref{fig:illustrate}).}


%% file: approach-compressed.tex
\section{Technical Design}\label{sec:approach}

\modify{This section describes the technical design of {\tech}. We start with an overview of {\tech} and then describe the details of each module.}

\subsection{Overview}
The overarching design goal of {\tech} is two-fold:
(1) \textit{large-scale}---being able to generate a massive number of samples to meet the thirsty need of training powerful defensive models, which requires the ability to inject vulnerabilities to 
normal code in the wild when following the injection-based methodology, and 
(2) \textit{high-quality}---not including too many noisy samples (i.e., those that are considered vulnerable but actually not) among the generated ones, without which the large scale would not be meaningful. With respect to the advances made by prior works and the challenges they face ($\S$\ref{sec:bkgmotive}), {\tech} achieves this goal \modify{with the design shown in Figure~\ref{fig:overview}. {\tech} consists of two main phases: \emph{learning/training} and \emph{production}. }

\modify{In the \emph{learning/training} phase, VGX gets itself built up through two main tasks: (1) train a deep learning based model that can identify the code contexts in a given (normal) program where vulnerabilities can be injected and (2) mine vulnerability-introducing code edit patterns that can materialize the injection in the identified contexts. These two tasks are achieved in two steps: \emph{semantics-aware contextualization} and \emph{human-knowledge-enhanced edit pattern formation}, followed by \emph{preprocessing} the datasets needed by this phase.}

\modify{In the \emph{production} phase, with the edit patterns and trained contextualization model, 
VGX starts with \emph{preprocessing} (using the same module) a given set of normal programs as input. Then, the model predicts potential vulnerability-injection code edit contexts, followed by matching and applying the most suitable edit patterns (for the contexts), resulting in (expectedly) vulnerable samples. }


\subsection{Preprocessing}
To enable \emph{semantics-aware contextualization} via the \emph{customized Transformer}, we use \emph{syntactic information} instead of processing code as natural language. 
Thus, 
we start by converting source code into Abstract Syntax Trees (ASTs), using 
tree-sitter~\cite{tree-sitter} as the AST parser. 
An AST is a tree composed of nodes, each of which has a node type, a set of child nodes, and a value if it is a terminal. 

Since Transformer takes text as inputs, we linearize the ASTs into text. 
As in~\cite{niu2022spt}, we do a pre-order traversal on the AST to get a sequence of node types and only keep the nodes above the expression level to reduce sequence length. 
We then concatenate the source code and the linearized AST for each sample with a special token {\tt [SEP]} between them. \modify{We use this mixed/dual input as it helps learn the syntactic and contextual structures of code~\cite{niu2022spt}}.


To make the \emph{customized Transformer} use \emph{semantic information}, which can be achieved via position encoding that is explicitly aware of value flow as discussed in $\S$\ref{sec:bkgmotive}, we construct a value flow graph (VFG) for each sample as in~\cite{guo2020graphcodebert}---because it can capture vulnerability related 
code semantics~\cite{guo2020graphcodebert,zhou2019devign}.
A VFG is a multi-edge graph $g(V,E)$, where V is a set of nodes each representing a variable and E is a set of edges that each represent the value flows between two variables. For instance, in an assignment statement 
{\tt c=a+b}, edges would exist from {\tt a} to {\tt c} and from {\tt b} to {\tt c}. The arrowed lines in Figure~\ref{fig:illustrate} show parts of the value flows in the motivating example. 

Our value-flow-based position encoding is based on the traditional one,  
where absolute position encoding is calculated as per the absolute position (i.e., index) of a given token. 
Yet we calculate this encoding of a variable based on its respective \emph{absolute VFG sub-graph}. To obtain this sub-graph, we traverse the VFG from the node of that variable until no new nodes can be found. For instance, in Figure~\ref{fig:illustrate}, the token {\tt password} at Line 9 has the value flow marked as red, and thus its absolute VFG sub-graph consists of itself and the node of token {\tt password} at Line 1.

\modify{Similarly, we calculates the relative position encoding for a pair of variables ($v1$,$v2$) based on their respective \emph{relative VFG sub-graph}.} To obtain this sub-graph, we traverse the VFG from the node of $v1$ until we reach the node of $v2$. 
For example, in Figure~\ref{fig:illustrate}, the pair of variables {\tt buf} at Line 5 and {\tt BUFSIZE} at Line 3 have their relative VFG sub-graph consist of the nodes of tokens {\tt buf} at Line 5, {\tt buf} at Line 3, and {\tt BUFSIZE} at Line 3.

\modify{Next, the resulting absolute and relative VFG sub-graphs will be used for position encoding as elaborated in $\S$\ref{ssec:pos-enc}}.

\subsection{Semantics-Aware Contextualization}\label{ssec:sac}

In this step, we train a {\tt customized Transformer} in order to achieve \emph{semantics-aware contextualization}. Given a normal program with the \emph{Code+AST} input, we expect the model to output a code fragment that can be manipulated to introduce a vulnerability, based on the semantics and context of the code. To that end, the {customized Transformer} leverages \emph{value-flow-based position encoding, pre-training,} and \emph{fine-tuning} with data augmentation. We separately describe these in the following subsections. 

\subsubsection{Value-flow-based Position Encoding}\label{ssec:pos-enc}
\modify{In the encoder of our {\tt customized Transformer}, the core module is self-attention blocks with position encoding:}

\vspace{-5pt}
\begin{equation}
    z_i=\sum^{n}_{j=1}\frac{exp(a_{ij})}{\sum^{n}_{j'=1}exp(a_{ij'})}(x_jW^V+r^{V}_{ij}+r^{V_{VFG}}_{ij})
\end{equation}
\vspace{-5pt}
\begin{multline}
    a_{ij}=\frac{1}{\sqrt{2d}}(x_iW^Q)(x_jW^K+r^{K}_{ij}+r^{K_{VFG}}_{ij})^T \\
    +\frac{1}{\sqrt{2d}}(a^Q_i)(a^K_j)^T+\frac{1}{\sqrt{2d}}(a_i^{Q_{VFG}})(a_j^{K_{VFG}})^T
\end{multline}

\modify{where $z_i$ is the output hidden representation for the $i$-th token, $a_{ij}$ is the attention between the $i$-th and $j$-th tokens; $x_i$ and $x_j$ are the hidden representation of the $i$-th and $j$-th tokens from the previous layer, respectively; $W^V$, $W^Q$, and $W^K$ are the weight matrix of \emph{value}, \emph{query}, and \emph{key}, respectively; $r^{V}_{ij}$, $r^{K}_{ij}$ are the traditional relative position encodings and $a^Q_i$ and $a^K_j$ are the traditional absolute position encodings; and $d$ is the dimension of the hidden representations.}
\modify{This traditional position encoding is solely based on the positions (indexes) of tokens hence \textit{lacks semantic understanding of code}. For example, a variable is defined at one line but used many lines after. In this case, the distance $(i-j)$ used by the traditional relative position encoding may be too large, although the two variable tokens may share a direct definition-use relationship and have a close semantic (i.e., value-flow) distance. }

\modify{To address this limitation, we utilize the VFG sub-graphs obtained during preprocessing to \textit{incorporate semantic information into the position encoding}. Specifically, we add our value-flow-based relative position encoding $r^{V_{VFG}}_{ij}$ and $r^{K_{VFG}}_{ij}$, as well as value-flow-based absolute position encoding $a_i^{Q_{VFG}}$, and $a_j^{K_{VFG}}$ to the self-attention block, where each of the position encoding is the graph encoding of the respective sub-graph. }

To compute the encoding of a given VFG sub-graph $g(V,E)$, we first use a pre-trained FastText~\cite{bojanowski2016enriching} model for C language to convert the variable name in each node into an embedding. Then, we use a  gated graph neural network (GGNN) to perform message passing aggregation and update the node embedding as follows:
\begin{equation}
    x'_v=GRU(x_v, \sum_{(u,v)\in E}g(x_u))
\end{equation}
where GRU is the gated recurrent function~\cite{chung2014empirical}, $g(.)$ is the function that assimilates the neighbor nodes' embedding, and $x_u$ is the neighbor node of $x_v$. \modify{Then, we sum up embeddings of the nodes in the graph to get the VFG sub-graph embedding. Finally, the graph embedding is multiplied with a weight matrix to get the \emph{value-flow-based position encoding} $r^{V_{VFG}}_{ij}$, $r^{K_{VFG}}_{ij}$, $a_i^{Q_{VFG}}$, and $a_j^{K_{VFG}}$}

Note that not all code tokens are variables. Thus, we only compute the value-flow-based position encoding when the token or both the tokens in the pair are variables. Otherwise, the value-flow-based position encoding is zero.

\subsubsection{Pre-Training}

To gain the awareness of code semantics, we performed pre-training before fine-tuning the model for \emph{semantics-aware contextualization}. We use three existing general programming language-oriented objectives from CodeT5~\cite{wang2021codet5}
and introduce two new code 
contextualization-specific objectives ({\tt CAP} and {\tt ISP}).

\textbf{CodeT5 Objectives.} 
We utilize three pre-training 
objectives from CodeT5~\cite{wang2021codet5} to 
learn general code comprehension 
for code-to-code transformation. The first 
is \emph{mask span prediction} ({\tt MSP}), where we randomly mask 15\% of the source code tokens in an input that contains both source code and AST, with each mask having a span length uniformly ranging from 1 to 5 tokens. We train the model to recover the masked tokens based on the context. The second objective is \emph{identifier tagging} ({\tt IT}), where the model is trained to predict whether each source code token is an identifier. The third objective is \emph{masked identifier prediction} ({\tt MIP}), where we mask all the identifiers in the source code, and the model is trained to recover the identifiers based on the code semantics. We follow the pre-training approach of CodeT5, and feed the three pre-training objectives alternatively during training with an equal probability.

\textbf{Code-AST Prediction (CAP).} Since our model takes source code and ASTs as input, similar to SPT-Code~\cite{niu2022spt}, we adopt their approach and pre-train our model with the Code-AST Prediction (CAP) objective. 
We assign the correct AST to the corresponding source code in 50\% of the pre-training samples, while in the other 50\% we randomly assign an incorrect AST. Our model is trained to predict if the assigned AST corresponds to the input source code.

\textbf{Irrelevant Statement Prediction (ISP).} 
For each pre-training sample, we insert at a random location a randomly chosen statement from another sample. \revise{In this case, the functionality of the sample is not impacted by removing the inserted statement. Thus, the inserted statement is \emph{semantically irrelevant} to the original sample. We pre-train the model to identify and output the irrelevant statement in each sample.}
We adopt this pre-training objective because it resembles our vulnerability injection localization objective. 
Since \textit{vulnerabilities are context-sensitive based on 
code semantics, learning to differentiate (semantically) irrelevant statements (from relevant ones) intuitively helps identify the right code context that is (semantically) relevant to the vulnerabilities to be injected}. 

We conduct pre-training in the following order: first against CAP, then the three CodeT5 objectives, and finally ISP, as justified by the inter-dependencies among these objectives. 
Also, for CAP and IT, the pre-training is done on the encoder only, while for all the other objectives (ISP, MSP, MIP) we pre-train both the encoder and decoder. 
These decisions are justified by the nature of the Transformer architecture. 

\subsubsection{Fine-Tuning}\label{ssec:ft}
After pre-training, we fine-tune the \emph{semantics-aware contextualization} model to locate code fragments that can be edited to introduce vulnerabilities, using the vulnerability-introducing code locations retrieved from the existing vulnerability fixes. To address the lack of fine-tuning data, we perform \emph{data augmentation} through code refactoring on the fine-tuning samples. Following the approach in \cite{nong2022generating}, we apply three types of \textit{semantics-preserving refactoring}: (1) reverse the condition in an {\tt if}-statement and swap the code in the {\tt if} and {\tt else} blocks; (2) convert a {\tt for} loop into an equivalent {\tt while} loop; and (3) insert unrelated junk code generated by SaBabi~\cite{sestili2018towards} to the code. We apply these transformations combinatorially on each original sample, leading to substantially more fine-tuning samples. 

\vspace{-0pt}
\subsection{Edit Pattern Formation}\label{ssec:hkeepf}
This section describes the process of {\tech} on learning (from the existing vulnerability fixes) to materialize the code editing for realizing vulnerability injection, including \textit{pattern extraction}, \textit{pattern filtering/ranking}, and \textit{pattern refinement and diversification}.
%

\subsubsection{Pattern Extraction}
We first parse the source code into ASTs like the one for \emph{semantics-aware contextualization} but with a different AST parser srcML~\cite{collard2013srcml}, as it better supports edit pattern extraction and application~\cite{bader2019getafix,nongvulgen}. Then, we follow the approach in Getafix~\cite{bader2019getafix} to extract the edit patterns via anti-unification. Because of the space limit, we refer readers to the original Getafix paper~\cite{bader2019getafix} for details. 
An edit pattern is a pair of code fragments in terms of ASTs representing a code edit. \modify{For example, "$h0[h1-1]=0; \Longrightarrow h0[h1]=0;$" represents removing "{\tt -1}" from the code where {\tt h0} and {\tt h1} are placeholders which can match any identifiers and literals.} After the pattern extraction, we get many edit patterns that range from very general (i.e., the patterns that can match and apply on many different code samples) to very specific (i.e., the patterns that can only match and apply on a few code samples)~\cite{bader2019getafix}. 

\begin{table}[tp]
  \centering
  \vspace{2pt}
  \caption{\modify{Manually Defined Vulnerability-Injection Patterns}}
  \vspace{-2pt}
    \scalebox{0.65}{
        \begin{tabular}{|p{40em}|}
        \hline
        \textbf{\ul{Patterns}:} *mutex*(h0); => EMPTY \\
        \textbf{\ul{Justification}:} 
        “Race Condition” mostly happens with a lack of mutex related statements, but there are many  mutex related function~\cite{race-condition}. Thus, once the located statement involve “mutex”, we delete it. \\
        \hline
        \textbf{\ul{Patterns}:} 
        *TCHECK*(h0); => EMPTY \quad
        *assert*(h0); => EMPTY \\
        \textbf{\ul{Justification}:} There are many samples in the training set deleting statements involving "TCHECK" and "assert", but they usually use different function names\cite{tcheck}. Thus, once located statement involve “TCHECK” or “assert”, we delete it. \\
        \hline
        \textbf{\ul{Patterns}:} *free*(h0); => EMPTY \qquad 
        *Free*(h0); => EMPTY \quad
        *destruct*(h0); => EMPTY \\
        \hspace{30pt}
        *destroy*(h0); => EMPTY \quad
        *unref*(h0); => EMPTY \qquad
        *clear*(h0); => EMPTY \\
        \textbf{\ul{Justification}:} "Memory Leak" mostly happens with not releasing assigned memory. However, there may be many different functions for releasing the memory~\cite{memleak}. Thus, once the the located statement involve memory release related functions, we delete it. \\
        \hline
        \textbf{\ul{Patterns}:} unsigned h0; => h0; \quad
        int64\_t h0; => int h0; \quad 
        static h0 h1 = h2; => h0 h1 = h2; \\
        \textbf{\ul{Justification}:} "Type Error" usually happens with not using static, unsigned, large-size types, but the current patterns specify too many details like identifier names or assigned values in the patterns~\cite{typeerr}. Thus, we make these details holes so that they are more general.  \\
        \hline
        \textbf{\ul{Patterns}:} 
        memset(h0); => EMPTY \quad
        h0 = *ERR*; => EMPTY \quad
        h0 = *NONE* => EMPTY \\
        \hspace{45pt}
        h0 = 0; => EMPTY \qquad
        h0 = NULL; => EMPTY \quad
        *buf* = h0; => EMPTY \\ 
        \textbf{\ul{Justification}:} "Use of Uninitialized Variables" usually happens with not initializing declared variables, but current patterns specify too many details like identifier names and values in the patterns~\cite{uninit}. Thus, we make these details holes and use regular expression to represent the common initialized value, so that they are more general. \\
        \hline
        \textbf{\ul{Patterns}:}      
        h0 = kcalloc(hole1, hole2, hole3);  => h0 = kzalloc(h1*h2, h3); \\
        \hspace{37pt}
        h0 = calloc(hole0, hole1); => h0 = malloc(h1*h2);\\
        \textbf{\ul{Justification}:} "Memory Allocation Vulnerability" usually happens when using unsafe memory allocation functions, but current patterns specify too many details like identifier names and values in the patterns~\cite{allocerr}. Thus, we make them holes to make the patterns more general. \\
        \hline
        \end{tabular}
    }
    \vspace{-5pt}
    \label{tab:derived-patterns}
\end{table}

\subsubsection{Pattern Filtering/Ranking}
To obtain the appropriate edit patterns for introducing vulnerabilities, we establish rules to filter and rank the extracted patterns. First,  we compute three scores for each of the extracted edit patterns:

(1) \textbf{Prevalence score ($s_{preval}$)}: \modify{Assuming knowing the vulnerabi-lity-introducing locations}, the prevalence score is the number of samples that can be injected the vulnerabilities correctly by this pattern in the training set, \modify{as it is proportional to the probability that the pattern can inject vulnerabilities successfully}.

(2) \textbf{Specialization score ($s_{spec}$)}: In the training samples, we compute the average number of AST subtrees that the pattern can match. The reciprocal of the average number is the specialization score, \modify{as it indicates whether the pattern is specific so that it will not match AST subtrees which cannot be injected vulnerabilities.}

(3) \textbf{Identifier score ($s_{ident}$)}: The identifier score is the number of specified identifier names in the pattern, \modify{as the identifier names inform about the code semantics significantly, which are crucial for correctly injecting vulnerabilities}.

\modify{The product of the three scores is the final ranking score.}
\vspace{-6pt}
\begin{equation}\label{rankingeq}
    s_{rank}=s_{preval}\times s_{spec}\times s_{ident} 
\vspace{-2pt}
\end{equation}

Then, we rank the edit patterns by 
$s_{rank}$
in a non-ascending order. 
The three individual scores are not normalized since only relative rankings matter.
The higher the score, the more likely the pattern can successfully inject vulnerabilities. 
Finally, we only keep the top 300 edit patterns, so as to filter out those that are not likely to inject vulnerabilities successfully.

\subsubsection{Pattern Refinement and Diversification}
\modify{While the patterns extracted and filtered/ranked 
enable to inject vulnerabilities in some cases, they 
do not fully capture the knowledge about how real-world vulnerabilities can be introduced to software as embodied in the extant vulnerability documentation such as CWE/CVEs in NVD and issues/bug reports on GitHub, because the available datasets of vulnerability-introducing edits are still limited, as discussed in $\S$\ref{sec:bkgmotive}.}
\modify{To address this limitation, we refine the patterns. We applied the 300 patterns to the vulnerability-introducing training samples, assuming knowing the vulnerability-introducing locations. As a result, we encountered \textit{false positives}, where the pattern was applied but the generated sample was not actually vulnerable, as well as \textit{false negatives}, where none of the 300 patterns could be applied.}

\modify{For the false positives, we consider that the patterns applied on them are too general. We removed the patterns against which more than half of the applications are false positives. In total, we removed 21 patterns in this manner.}

\modify{For the false negatives, we consider that the corresponding patterns are too specific. Thus, we manually define new patterns that are necessarily more general. 
Specifically, 
we examined the vulnerability-fixing commit 
associated with the false-negative training sample 
and identified the corresponding CWE ID~\cite{cwereport}. 
Then, we read the CWE documentation and check the examples used to describe that CWE. We further examined other real-world vulnerability samples with the same CWE ID in the NVD/CVE database~\cite{nvd}.  
Based on the false negative, the examples, and other real-world samples, we compose 
several possible patterns that would be able to inject the respective vulnerabilities. Then, we went back to the existing pattern set and found patterns close to the composed ones. We finally modified those patterns into the composed ones with the use of regular expression (regex). These modified patterns are the 20 manually defined, new patterns described in Table~\ref{tab:derived-patterns}.
}

\begin{table}[tp]
  \centering
  \vspace{-0pt}
  \caption{\modify{Pattern Mutation Rules}}
  \vspace{-10pt}
    \scalebox{0.65}{
    \begin{tabular}{|p{38em}|}
    \hline
    \textbf{\ul{Rule}:} function\_name(parameters); =>new\_function\_name(new\_parameters); {\color{red}$\boldsymbol{\Leftrightarrow}$} \\
    \hspace{24pt}h0=function\_name(parameters); => h0=new\_function\_name(new\_parameters);\\
    \textbf{\ul{Justification}:} 
    Some function calls like strncpy may return some values but the return values do not always assign to a variable. Mutating such patterns so that they have or remove the return value assignments increases the generalizability of the pattern set.\\
    \hline
    \textbf{\ul{Rule}:} if(condition) {return NULL/0/-1;} => EMPTY {\color{red}$\boldsymbol{\Leftrightarrow}$}\\
    if(condition) {return -EINVAL/EBADFD/ENOTSOCK/EPERM/ENODEV/ENOMEM;} => EMPTY \\
    \textbf{\ul{Justification}:} 
    The typical safety issue checks do some check in an if statement condition and then return  an error code when found the issue. However,  there are many possible returned error codes. We mutate these returned error codes to make the patterns more general. \\
    \hline
    \textbf{\ul{Rule}:} 
    if(specific\_condition) {return error\_code;} => EMPTY 
    {\color{red}$\boldsymbol{\Rightarrow}$} \\
    \hspace{24pt}if(hole) {return error\_code;}=> EMPTY \\
    \textbf{\ul{Justification}:} In our automatic pattern mining, the mined safety checks if statements may be too specific in the condition. However, if an if statement only has one return statement, it is very likely that it is a safety check if statement. Thus, we remove the specific conditions in these if statements and use a hole to make the patterns more general. \\
    \hline
    \textbf{\ul{Rule}:} 
    if(condition) {return error\_code;} => EMPTY 
   {\color{red}$\boldsymbol{\Leftrightarrow}$}
    if(condition) {break/continue;} => EMPTY \\
    \textbf{\ul{Justification}:} When the safety checks if statements find issues, they may not always exit the function using a return. If the safety check is in a for/while/switch block, it may use a break or continue to exit. Thus, we mutate the exit statement to make the patterns more general. \\
    \hline
    \end{tabular}}
    \vspace{-6pt}
  \label{tab:mutation}
\end{table}


Since manual refinement is costly, we are not able to derive many new patterns. 
Some remaining patterns are still too specific in syntactic structure. Thus, we further derived a set of pattern mutation rules to make the edit patterns 
sufficiently
general. \modify{We carefully checked the traditional mutation operators for C language~\cite{agrawal1989design}. We then selected those that do not change the code functionality significantly and finally derived four types of pattern mutation rules. Table~\ref{tab:mutation} shows the pattern mutation rules we derived and the respective justifications for them. The red $\Leftrightarrow$ means that the mutation is bidirectional, while $\Rightarrow$ means that the mutation is unidirectional.} After the mutation, we obtained a total of 604 high-quality \emph{vulnerability-injection patterns}.


\revise{Currently, the pattern refinement and pattern mutation-rule derivation are carried out manually and aimed at the C language. Yet the process can be readily reproduced and replicated by following the principled steps laid out above. This also largely makes our {\tech} approach adaptable for other programming languages.}

\vspace{-0pt}
\subsection{Vulnerability Production}\label{ssec:production}
With the trained \emph{contextualization} model and the \emph{vulnerability-injection code edit patterns}, {\tech} is ready to produce vulnerability data. Given a normal program, we again preprocess it like we do in the learning/training phase. Next, the 
\emph{contextualization} model 
identifies 
a code fragment (as the context) for vulnerability injection. Then, {\tech} applies the most suitable pattern to the 
identified context. 
To do so, we rank the patterns in the final set of \emph{vulnerability-injection code edit patterns} 
by the score computed in Equation~\ref{rankingeq} 
and apply the first 
pattern in the set that can match the 
context. 
Note that {\tech} only injects the vulnerability when one of the patterns matches the context. 
If none of the patterns matches, {\tech} discards the sample to reduce false positives. Otherwise, the program with the pattern applied is expected to be vulnerable.

%% file: evaluation.tex
\vspace{-0pt}
\section{Evaluation}
We evaluate the effectiveness of {\tech} for generating vulnerability data. We seek to answer the following research questions:
\begin{itemize}[leftmargin=*]
    \item \textbf{RQ1:} How effective is {\tech} in vulnerability generation compared to other approaches?
    \item \textbf{RQ2:} How do the novel design components contribute?
    \item \modify{\textbf{RQ3:} How efficient is {\tech} in vulnerability generation?}
\end{itemize}


\vspace{-0pt}
\subsection{Implementation}

\revise{To ensure the accuracy/reliability of the value flow graphs created in the preprocessing module, we leverage the value flow extraction tool from GraphCodeBERT~\cite{guo2020graphcodebert} to construct the value flow graph of each sample.}
We modify the source code of TPTrans~\cite{peng2021integrating} to build our \emph{customized Transformer} \modify{as it provides position encoding implementation which is easy to adapt/customize}. We use the Getafix implementation in VulGen~\cite{nongvulgen} for the anti-unification-based pattern extraction. Our experiments were performed on a machine with a 32 Cores AMD Ryzen 3970X (3.7GHz) CPU, two Nvidia RTX 3090 GPUs, and 256GB DDR memory.

\vspace{-0pt}
\subsection{Dataset}
For the \emph{pre-training} in \emph{Semantics-Aware Contextualization}, we use the IBM CodeNet~\cite{puri2021codenet} dataset.
We extract 1,213,907 functions in C language from it to build the pre-training dataset.
\modify{For the \emph{fine-tuning} in \emph{Semantics-Aware Contextualization} and \emph{Edit Pattern Formation}, we use the dataset in VulGen~\cite{nongvulgen} which is the combination of five widely used vulnerability 
datasets. We remove the overlapped samples and eventually obtain 7,764 samples for the evaluation. We split the dataset by 9:1 for training (6,989) and testing (775). We then augment ($\S$\ref{ssec:ft}) the 6,989 samples into 156,665 for \emph{fine-tuning}.}

\vspace{-0pt}
\subsection{Metrics}\label{ssec:metrics}
\modify{Given that a high-quality vulnerability sample should be a truly exploitable program~\cite{croft2023data}, we evaluate whether the code changes make the program exploitable for the attackers. For the samples exactly-matching their ground truths, they are known as exploitable because they are confirmed real-world vulnerabilities. 
For these cases, we evaluate vulnerability generation in terms of precision, recall, and F1.}
We compute: 
$precision=\frac{\#matched\ samples}{\#generated\ samples}$; $recall=\frac{\#matched\ samples}{\#testing\ samples}$ and $F1=\frac{2\times recall\times precision}{precision+recall}$. 

Since the generated samples may be exploitable although do not exactly match the ground truth~\cite{nong2022generating,nongvulgen}, we randomly sampled and manually inspected some of them to compute the $success\ rate=\frac{\#exploitable\ samples}{\#generated\ samples}$ where the $exploitable$ $samples$ 
also include the $matched$ $samples$. 
We inspected 258 cases, a sample size that is statistically significant
at 95\% confidence level and 5\% margin of error with respect to the 
population (i.e., testing set) size of 775. 

\modify{The manual inspection of a sample assesses if an exploit can be written to attack it. 
Yet manually writing exploits 
for all the non-exactly-matching samples generated by {\tech} and baseline techniques  
is quite difficult.
Thus, we wrote exploits for 30\% of 
those samples in RQ1 and RQ4. 
Each exploit is an executable 
that can attack the generated program but not the normal one,  
ensuring the exploitability is introduced by the code change. 
This process 
helps us learn how to decide whether a 
sample is exploitable, based on which we determined the exploitability of 
the remaining 70\% without actually writing/running the exploits. 
The 
inspection was first done by the first author, checked by the third and fourth authors, and then verified/calibrated by all three to ensure the correctness.}

\subsection{RQ1: Effectiveness of {\tech}}\label{ssec:effectiveness}

We assess the effectiveness of {\tech} over four baselines: 
\vspace{-0pt}
\begin{itemize}[leftmargin=*]
    \item \textbf{VulGen}~\cite{nongvulgen} is a vulnerability generation tool which fine-tunes CodeT5~\cite{wang2021codet5} for localization and uses Getafix~\cite{bader2019getafix} to mine edit patterns for vulnerability injection.
    \item \textbf{CodeT5}~\cite{wang2021codet5} is a pre-trained model good at code-to-code transformation, which can be directly fine-tuned to generate vulnerable code for given normal programs.
    \item \textbf{Getafix}~\cite{bader2019getafix} is a pattern mining and application technique for bug fixing. We 
    directly use it 
    for vulnerability generation.
    \item \textbf{Graph2Edit}~\cite{yao2021learning} is a code editing model taking a program's AST as input and predicting a sequence of edits on it, which can be trained for vulnerability injection.
\end{itemize}
\vspace{0pt}

Table~\ref{tab:result} shows the effectiveness of {\tech} and the baselines against the 775 testing samples. {\tech} generates 296 samples, of which 176 exactly match the ground truth. Thus, the precision, recall, and F1 is 59.46\%, 22.71\%, and 32.87\%, respectively. The high precision indicates 
the good quality 
of the generated samples, and the 22.71\% recall 
suggests the generalizability 
for supporting 
large-scale vulnerability 
generation. The 32.87\% F1 
shows 
the overall promising effectiveness of {\tech}---99\% higher than the best baseline. 
Notably, Column 3 shows {\tech}'s success rate of \modify{93.02\%}, indicating that the vast majority of its generated samples are vulnerable/exploitable. 


%

Table~\ref{tab:result} Rows 3-6 show the effectiveness of the four baselines, and the numbers in parentheses indicate {\tech}'s relative improvements over the baselines. {\tech} outperforms the GNN-based code editor Graph2Edit by substantial margins.  
While Graph2Edit is a general-purpose code editor with a promising editing process (i.e., predicting a sequence edits on the AST)~\cite{nong2022generating}, it suffers from the lack of training samples as GNN needs a large number of samples to be reasonably trained~\cite{nongvulgen}. \modify{{\tech} outperforms the CodeT5 and Getafix also quite significantly. 
} 
This may be justified by the merits of combining deep learning to locate injection with a pattern-based method to materialize injection edits. While VulGen also works in an overall similar way, {\tech} still greatly outperforms it. 
This is due to {\tech} taking the \modify{\emph{advantages of human-knowledge-enhanced edit patterns and semantics-aware contextualization, which largely overcome VulGen's Limitation \textcircled{1} and Limitation \textcircled{2}, respectively.}}

\begin{table}[t]
\center
\vspace{0pt}
\caption{
Effectiveness (and improvements in parentheses) of {\tech} over the baselines for vulnerability generation in RQ1.
}
\vspace{-0pt}
\scalebox{0.65}{
\begin{tabular}{lllll}
    \hline
    \textbf{Technique} & \textbf{Precision} & \textbf{Recall} & \textbf{F1} & \textbf{Success Rate} \\
    \hline
    \textbf{\tech} & 59.46\% & 22.71\% & 
    \textbf{32.87\%} & \textbf{\modify{93.02\%}}\\
    \textbf{VulGen} & 17.50\% (239.77\%$\uparrow$) & 15.74\% (44.28\%$\uparrow$) & \textbf{16.51\% (99.09\%$\uparrow$)} & \textbf{\modify{75.96\% (22.45\%$\uparrow$)}}\\
    \textbf{CodeT5} & 12.65\% (370.04\%$\uparrow$) & 12.65\% (79.53\%$\uparrow$) & \textbf{12.65\% (159.84\%$\uparrow$)} & \textbf{\modify{24.81\% (274.93\%$\uparrow$)}} \\
    \textbf{Getafix} & 4.67\% (1173.23\%$\uparrow$) & 2.58\% (780.23\%$\uparrow$) & \textbf{3.32\% (890.06\%$\uparrow$)} & \textbf{\modify{57.75\% (67.07\%$\uparrow$)}} \\
    \textbf{Graph2Edit} & 13.97\% (325.62\%$\uparrow$) & 13.97\% (65.56\%$\uparrow$) & \textbf{13.97\% (135.29\%$\uparrow$)} & \textbf{\modify{21.71\% (328.47\%$\uparrow$)}}\\
    \hline
\end{tabular}}
\vspace{-0pt}
\label{tab:result}
\end{table}

\subsection{RQ2: Contributions of Novel Components}
In this section, we investigate the contribution of each novel component in {\tech} through ablation studies. 
We remove each of 
those components 
and compare the effectiveness before and after the removal. The results are shown in Table~\ref{tab:ablation}. To show the impacts of the contextualization designs, we also report the localization accuracy in Column 2. The numbers in parentheses indicate {\tech}'s relative improvements compared to the 
ablated versions. 


To assess the impact of the \emph{contextualization-specific pre-training objectives}, we remove the pre-training for {\tt CAP} and {\tt ISP}. Table~\ref{tab:ablation} Row 3 shows the results. With the two pre-training tasks, {\tech} improves the localization accuracy by 4.37\%, and thus improves the vulnerability generation by 9.38\%, 8.55\%, and 8.87\% in terms of precision, recall, and F1, respectively. This indicates that the contextualization-specific objectives are effective for pre-training.

To understand the contribution of the \emph{linearized AST}, we remove the AST part in the Transformer input and only use the source code. Table~\ref{tab:ablation} Row 4 shows the results. With linearized AST, {\tech} improves the localization accuracy by 11.44\%, and thus improves the vulnerability generation effectiveness by 10.32\%, 15.81\%, and 14.29\% in terms of precision, recall, and F1, respectively. This indicates that adding the linearized AST to the input text effectively helps the model understand the syntactic structure of the code, 
hence notably improving vulnerability injection. 

\begin{table}[t]
\center
\caption{The contribution of {\tech}'s components for vulnerability generation. The numbers in parentheses are {\tech}'s relative improvements compared to the ablated versions.} 
\vspace{-0pt}
\scalebox{0.62}{
\begin{tabular}{lllll}
    \hline
    \textbf{Experiment} & \textbf{Loc Acc} & \textbf{Precision} & \textbf{Recall} & \textbf{F1} \\
    \hline
    \textbf{\tech} & 55.35\% & 59.46\% & 22.71\% & 
    \textbf{32.87\%}\\
    \textbf{No CAP and ISP} & 53.03\% (4.37\%$\uparrow$) & 54.36\% (9.38\%$\uparrow$) & 20.90\% (8.66\%$\uparrow$) & \textbf{30.19\% (8.87\%$\uparrow$)} \\
    \textbf{No AST} & 49.67\% (11.44\%$\uparrow$) & 53.90\% (10.32\%$\uparrow$) & 19.61\% (15.81\%$\uparrow$) & \textbf{28.76\% (14.29\%$\uparrow$)}  \\
    \textbf{No VFG} & 52.13\% (6.18\%$\uparrow$) & 53.29\% (11.58\%$\uparrow$) & 21.93\% (3.56\%$\uparrow$) & \textbf{31.07\% (5.79\%$\uparrow$)}  \\
    \textbf{No Augmentation} & 51.61\% (7.25\%$\uparrow$) & 53.33\% (11.49\%$\uparrow$) & 19.61\% (15.81\%$\uparrow$) & \textbf{28.67\% (14.65\%$\uparrow$)} \\
    \textbf{No Diversification} & 55.35\% (0.00\%$\uparrow$) & 62.45\% (-4.78\%$\uparrow$) & 20.38\% (11.43\%$\uparrow$) & \textbf{30.73\% (6.96\%$\uparrow$)} \\
    \makecell[l]{\textbf{No Refinement}}\ & 55.35\% (0.00\%$\uparrow$) & 71.91\% (-17.31\%$\uparrow$) & 16.51\% (37.55\%$\uparrow$) & \textbf{26.85\% (22.42\%$\uparrow$)} \\
    
    \hline
\end{tabular}}
\vspace{-0pt}
\label{tab:ablation}
\end{table}

To show the effectiveness of the \emph{VFG-based position encoding}, we remove it and only use traditional position encoding in the Transformer model. Table~\ref{tab:ablation} Row 5 shows the results. With our new position encoding, {\tech} improves the localization accuracy by 6.18\%, and thus improves the vulnerability generation effectiveness by 11.58\%, 3.56\%, and 5.79\% in terms of precision, recall, and F1, respectively. This indicates that the VFG-based position encoding helps the Transformer understand the semantics of the code, thus improves the vulnerability generation significantly.

To measure the impact of the \emph{data augmentation when fine-tuning} the contextualization model, we use the original 6,989 training samples only to train it. Table~\ref{tab:ablation} Row 6 shows the results. With the data augmentation, The localization accuracy improves by 7.25\%, and thus improves the vulnerability generation effectiveness by 11.49\%, 15.81\%, and 14.65\% in terms of precision, recall, and F1, respectively. This indicates that the data augmentation mitigates the lack of training data, thus is helpful for vulnerability generation.

To see the impact of \emph{pattern diversification}, we only use the edit patterns prior to the diversification (mutation). Table~\ref{tab:ablation} Row 7 shows the results. While pattern diversification makes the precision decrease by 4.78\%, the recall and F1 improve by 11.43\% and 6.96\%, respectively. While the mutation makes the vulnerability injection more general, it also makes the patterns less restrictive against the injection context identified. The improvement of recall indicates that {\tech} is able to generate more vulnerable samples, which is important for generating large-scale vulnerability datasets.

We further rollback the edit pattern set to the one without \emph{pattern refinement}. Table~\ref{tab:ablation} Row 8 shows the results. We can see the recall and F1 further decrease to 16.51\%, and 26.85\%, respectively, while the precision increases to 71.91\%. This indicates that patterns without refinement are too strict and specific, making it difficult for {\tech} to generate large-scale vulnerability data. Thus, our refinement improves the generalizability of the patterns, making the overall vulnerability generation 
more scalable. 

\subsection{\modify{RQ3: Efficiency of {\tech}}}
\modify{
As VGX, 
VulGen, and Getafix 
support multiprocessing generation but 
CodeT5 and Graph2Edit do not, 
we use single-process generation for a fair comparison. On average, {\tech} injects vulnerabilities on 189.02 samples per minute, while the numbers for VulGen, Getafix CodeT5, and Graph2Edit are 132.85, 276.78, 29.88, and 89.08, indicating {\tech} is comparable to the baselines in terms of efficiency.}

\section{Usefulness of {\tech}} \label{sec:usefulness}
We use {\tech} to generate a large-scale vulnerability dataset and evaluate the usefulness of the data for training downstream DL-based vulnerability analysis tools. \modify{As VulGen is also designed for vulnerability data generation and is shown to be relatively effective in the original paper~\cite{nongvulgen}, we also directly compare {\tech}'s usefulness with VulGen.} We answer the following research questions:
\begin{itemize}[leftmargin=*]
    \item \textbf{RQ4}: How effective is {\tech} in generating  large-scale vulnerable data using normal programs in the wild?
    \item \textbf{RQ5}: Can the generated samples improve the downstream DL-based vulnerability analysis tools?
    \item \textbf{RQ6}: Can the improved vulnerability detection models find more latest real-world vulnerabilities?
\end{itemize}

\subsection{RQ4: Large-Scale Production}\label{ssec:wild}
To get a large number of normal samples for {\tech} to generate vulnerability data, we follow the approach in~\cite{fan2020ac} to collect the latest source code of 238 projects that are involved in the CVE/NVD database~\cite{nvd}. Then, we extract the functions in these projects and obtain 738,453 normal functions for vulnerability injection. 

We apply {\tech} on these functions and it generates \textbf{150,392 samples} in 50 hours 48 minutes. We randomly sample some to manually inspect. With the same procedure in $\S$\ref{ssec:metrics}, the sample size is 375. \modify{Again, we write exploits for 30\% (102) of the samples and then directly label the remaining samples. The success rate is \textbf{90.13\%}, which is close to the success rate (93.03\%) 
for RQ1 ($\S$\ref{ssec:effectiveness}).}
\modify{In comparison, VulGen generates \textbf{686,513 samples}. With the sample size 384, the success rate is \textbf{41.33\%}. This is not only much lower than 
{\tech}'s, but also much lower than its own (75.96\%) in $\S$\ref{ssec:effectiveness}. The reason is that 
VulGen's edit patterns are vague hence applied on the statements that cannot be injected vulnerabilities (Limitation \textcircled{1}) and the localization model is not semantics-aware hence less accurate (Limitation \textcircled{2}). This shows that {\tech} is effective to generate large-scale vulnerability data from the wild while VulGen is not.}

\begin{figure}[tp]
	\vspace{-0pt}
	\includegraphics[width=1.0\linewidth]{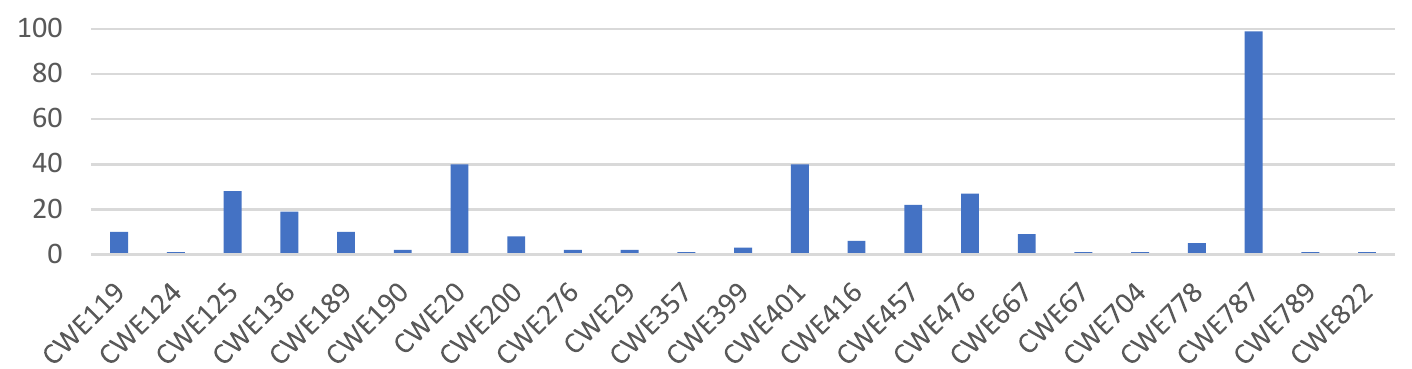}
        \vspace{-14pt}
	\caption{\modify{The vulnerability type (CWE) distribution of generated vulnerable samples in the large-scale production.}}
	\label{fig:vultype}
	\vspace{-0pt}
\end{figure}

We also checked the vulnerability types represented by 
these \ul{375
samples} and found that they cover \modify{\ul{23 
types (CWEs)}. Figure~\ref{fig:vultype}} shows the distribution of them. It is worth noting that 
these types include 
many of the most (e.g., top-25) dangerous ones 
according to the recent CWE report~\cite{cwereport}, such as CWE-787, CWE-20, CWE-125, etc. This indicates that our generated vulnerability samples are diverse. 


\subsection{RQ5: Downstream Analysis Improvement}\label{ssec:downstream}
Given the vulnerability generation process, our data supports model training for at least three downstream tasks: (1) \emph{function-level vulnerability detection}; (2) \emph{line-level vulnerability localization}; and (3) \emph{vulnerability repair}. We use the generated data to augment the training sets of the models and show the improvement after the augmentation. Since some models are designed for the existing small vulnerability datasets, they are not scalable for our large dataset in terms of the time cost and the memory usage. Thus, \modify{we use 10\% of {\tech}'s generated samples (15,039) and VulGen's generated samples (68,651) to improve the downstream models. To show that {\tech}'s generated samples are more practical than the existing synthetic samples, we also use the same number (15,039) of samples from the widely used synthetic dataset SARD~\cite{black2017sard} to improve the downstream tasks for comparison.} Since the original testing sets of these models are split from their training sets which have the same distribution, testing the models on them may not show their real performance~\cite{nong2022open}. Thus, we follow the approaches in \cite{nongvulgen,nong2022generating} to leverage the \emph{independent testing} setting: for each downstream task, we use a third-party testing set that is from a different source from the original training sets. \modify{Again, we ensure that there is no overlapping between the training sets and testing sets.}

\begin{table}[t]
\center
\caption{\modify{Improvement of vulnerability detection using {\tech}'s, VulGen's generated samples, and samples from SARD.}}
\vspace{-0pt}
\scalebox{0.7}{
\begin{tabular}{llll}
    \hline
    \textbf{Model} & \textbf{Precision} & \textbf{Recall} & \textbf{F1} \\
    \hline
    \textbf{Devign-ori} & 9.82\% & 50.19\% & 
    \textbf{16.43\%} \\
    \textbf{Devign-aug-VGX} & 12.37\% (25.97\%$\uparrow$) & 52.47\% (4.54\%$\uparrow$) & 
    \textbf{20.01\% (21.79\%$\uparrow$)} \\ 
    \textbf{Devign-aug-VulGen} & 11.23\% (14.35\%$\uparrow$) & 30.03\% (-40.17\%$\uparrow$) & 
    \textbf{16.35\% (-0.49\%$\uparrow$)} \\ 
    \textbf{Devign-aug-SARD} & 15.27\% (55.49\%$\uparrow$) & 15.21\% (-69.69\%$\uparrow$) & 
    \textbf{15.24\% (-7.24\%$\uparrow$)} \\ 
    \hline
    \textbf{LineVul-ori} & 26.42\% & 2.52\% & 
    \textbf{4.61\%} \\
    \textbf{LineVul-aug-VGX} & 11.38\% (-56.93\%$\uparrow$) & 78.00\% (2995\%$\uparrow$) & 
    \textbf{19.86\% (330.80\%$\uparrow$)} \\ 
    \textbf{LineVul-aug-VulGen} & 9.97\% (-62.26\%$\uparrow$) & 3.73\% (48.01\%$\uparrow$) & 
    \textbf{5.42\% (17.57\%$\uparrow$)} \\ 
    \textbf{LineVul-aug-SARD} & 9.19\% (-65.21\%$\uparrow$) & 85.70\% (3300\%$\uparrow$) & 
    \textbf{16.60\% (260.09\%$\uparrow$)} \\ 
    \hline
    \textbf{IVDetect-ori} & 9.06\% & 75.52\% & 
    \textbf{16.18\%} \\
    \textbf{IVDetect-aug-VGX} & 13.21\% (45.81\%$\uparrow$) & 35.66\% (-52.78\%$\uparrow$) & 
    \textbf{19.28\% (19.15\%$\uparrow$)} \\
    \textbf{IVDetect-aug-VulGen} & 7.90\% (-12.80\%$\uparrow$) & 65.03\% (-13.89\%$\uparrow$) & 
    \textbf{14.09\% (-12.92\%$\uparrow$)} \\
    \textbf{IVDetect-aug-SARD} & 10.04\% (10.81\%$\uparrow$) & 55.94\% (-25.92\%$\uparrow$) & 
    \textbf{17.02\% (5.19\%$\uparrow$)} \\
    \hline
\end{tabular}}
\vspace{-0pt}
\label{tab:res-detect}
\end{table}

\subsubsection{Function-Level Vulnerability Detection}\label{ssec:detect}
We improve three vulnerability detectors, Devign~\cite{zhou2019devign}, LineVul~\cite{fu2022linevul}, and IVDetect~\cite{li2021vulnerability}, as they are the SOTA tools publicly available for replication experiments. The original training set of each tool has vulnerable and non-vulnerable samples. To improve each training set, we add the 15,039 generated samples by {\tech} and the \textit{same proportion} of non-vulnerable samples from our normal sample set used to generate vulnerable data. Specifically, the training set of Devign has 9,744 vulnerable and 11,012 non-vulnerable samples, thus we add {\tech}'s 15,039 generated samples and 16,996 non-vulnerable samples to the training set. 
Both LineVul and IVDetect use the Fan~\cite{fan2020ac} training set with 10,547 vulnerable and 168,752 non-vulnerable samples, thus we add {\tech}'s 15,039 generated samples and 240,624 non-vulnerable samples to the training set to improve the models.

To leverage \textit{independent testing}, we test the models on the ReVeal dataset~\cite{chakraborty2022deep}. It has 1,664 vulnerable and 16,505 non-vulnerable samples which are manually collected from real-world projects. 
Table~\ref{tab:res-detect} shows the results of the trained models of the three tools, where \textit{-ori} means that the model is trained on the original training set and \modify{\textit{-aug-{\tech}} means that the model is trained on the augmented training set using {\tech}'s generated samples}. \revise{We use the most widely adopted metrics for the vulnerability detection models, which are precision, recall, and F1, to assess the improvement}. We notice that the F1 scores, which represent the overall performance of the tools, improve by 21.79\%, 330.80\%, and 19.15\% for Devign, LineVul, and IVDetect, respectively. This indicates that {\tech}'s generated data is able to improve vulnerability detection significantly. 

\modify{We also do the same process with VulGen's generated samples and SARD samples. We notice that the improvements are much lower than the one with {\tech}'s generated samples (e.g., 17.57\% and 260.09\% versus 330.80\% F1 against LineVul) and they may even decrease the performance (e.g., 0.49\% and 7.24\% F1 decrease against Devign). Besides, VulGen's generated samples may be even worse than the SARD samples. The reason is that VulGen's generated samples have a low success rate (41.33\%) for large-scale production and the label noise weakens the data augmentation seriously.}

\begin{table}[t]
\center
\caption{\modify{Improvement of vulnerability localization using {\tech}'s, VulGen's generated samples, and samples from SARD.}}
\vspace{-10pt}
\scalebox{0.65}{
\begin{tabular}{ll|ll}
    \hline
    \textbf{Model} & \textbf{Top-10 Accuracy} & \textbf{Model} & \textbf{Top-10 Accuracy}  \\
    \hline
    \textbf{LineVul-ori} & 48.84\% & \textbf{LineVD-ori} & 59.25\%  \\
    \textbf{LineVul-aug-VGX} & 58.27\% (19.31\%$\uparrow$)  &  \textbf{LineVD-aug-VGX} & 66.87\% (12.86\%$\uparrow$)  \\ 
    \textbf{LineVul-aug-VulGen} & 53.43\% (9.39\%$\uparrow$) & \textbf{LineVD-aug-VulGen} & 52.68\% (-11.09\%$\uparrow$)  \\ 
    \textbf{LineVul-aug-SARD} & 49.85\% (2.07\%$\uparrow$)  & \textbf{LineVD-aug-SARD} & 64.18\% (8.32\%$\uparrow$)  \\
    \hline
\end{tabular}}
\vspace{-0pt}
\label{tab:res-loc}
\end{table}

\begin{table}[t]
\center
\caption{\modify{Improvement of vulnerability repair using {\tech}'s, VulGen's generated samples, and samples from SARD.}}
\vspace{-10pt}
\scalebox{0.65}{
\begin{tabular}{llll}
    \hline
    \textbf{Model} & \textbf{Top-1 Accuracy} & \textbf{Top-5 Accuracy} & \textbf{Top-50 Accuracy} \\
    \hline
    \textbf{VulRepair-ori} & 8.55\% & 11.81\% & 16.29\%   \\
    \textbf{VulRepair-aug-VGX} & 21.05\% (146.20\%$\uparrow$) & 29.12\% (146.57\%$\uparrow$) & 30.14\% (85.02\%$\uparrow$)    \\
    \textbf{VulRepair-aug-VulGen} & 11.81\% (38.13\%$\uparrow$) & 16.77\% (41.20\%$\uparrow$) & 17.85\% (9.85\%$\uparrow$)    \\
    \textbf{VulRepair-aug-SARD} & 11.07\%  (29.47\%$\uparrow$)& 13.92\% (17.87\%$\uparrow$) & 17.18\% (5.46\%$\uparrow$)     \\ \hline
    \textbf{VRepair-ori} & 2.58\% & 5.16\% & 8.62\% \\
    \textbf{VRepair-aug-VGX} & 4.41\% (70.93\%$\uparrow$) & 10.59\% (105.23\%$\uparrow$) & 17.18\% (99.30\%$\uparrow$) \\
    \textbf{VRepair-aug-VulGen} & 2.85\% (10.46\%$\uparrow$) & 7.26\% (40.70\%$\uparrow$) & 14.05\%  (62.99\%$\uparrow$) \\
    \textbf{VRepair-aug-SARD} & 1.36\% (-47.28\%$\uparrow$) & 3.46\% (-32.94\%$\uparrow$) & 4.96\% (-42.45\%$\uparrow$) \\
    \hline
\end{tabular}}
\vspace{-0pt}
\label{tab:res-repair}
\end{table}

\subsubsection{Line-Level Vulnerability Localization.} We improve two vulnerability localization tools LineVul~\cite{fu2022linevul} and LineVD~\cite{hin2022linevd} with {\tech}'s generated samples. Their training set 
is also Fan~\cite{fan2020ac} mentioned in $\S$\ref{ssec:detect}. Thus, we use the same augmentation setting. We again test the models on the ReVeal dataset because it provides the vulnerable lines for the 1,664 vulnerable samples. Table~\ref{tab:res-loc} shows the results of before and after the improvement. \revise{We adopt the commonly used metric in LineVul and LineVD, which is top-10 accuracy, to evaluate the improvement. }
The improvements brought by {\tech}'s generated samples are 19.31\% and 12.86\% for LineVul and LineVD, respectively. This indicates that the samples are useful for improving vulnerability localization. \modify{In comparison, VulGen's generated samples and SARD samples have lower improvements compared to {\tech}'s generated samples or even decrease performance (9.39\% and 2.07\% improvements for LineVul,  respectively; and -11.09\% and 8.32\% improvements for LineVD, respectively).}

\subsubsection{Vulnerability Repair}
We improve the latest two vulnerability repair tools VulRepair~\cite{fu2022vulrepair} and VRepair~\cite{chen2022neural}. Both use the vulnerability repair samples from the combination of Fan~\cite{fan2020ac} and CVEFixes~\cite{bhandari2021cvefixes}, which consist of 8,482 pairs of vulnerable and the respective non-vulnerable programs, to train the models. We thus use the 15,039 pairs of {\tech}'s generated vulnerable and the respective non-vulnerable programs to augment the training set.
We test the models on the PatchDB dataset~\cite{wang2021patchdb} because its vulnerability repair samples are from real-world projects and confirmed by humans. Table~\ref{tab:res-repair} shows improvement results. \revise{We adopt the commonly used settings where the beam search sizes are 1, 5, and 50 and thus they report top-1, top-5, top-50 accuracy. With {\tech}'s generated samples, the top-1, top-5, and top-50 accuracy of VulRepair improves by 146.20\%, 146.57\%, and 85.02\%, respectively. The top-1, top-5, and top-50 accuracy of VRepair improves by 70.93\%, 105.23\%, and 99.30\%, respectively.
This indicates that the samples are quite useful to boost vulnerability repair. In comparison, VulGen's generated samples and SARD samples bring lower improvements 
or even decrease the performance (e.g., 9.85\% and 5.46\% improvements for VulRepair's Top-50 accuracy, respectively; and 62.99\% and -42.45\% for VRepair's Top-50 accuracy, respectively).}
\begin{table}[t]
  \centering
  \caption{Latest CVEs Detected by Improved LineVul but missed by the original one.}
  \vspace{-10pt}
    \scalebox{0.65}{
    \begin{tabular}{lcc|lcc}
    \hline
    \textbf{CVE ID}   & \textbf{Project} & \textbf{CWE ID} & \textbf{CVE-ID}   & \textbf{Project} & \textbf{CWE ID} \\ \hline
    2022-46149 & Cap'n Proto & CWE-125 & 2021-3764 & Linux Kernel & CWE-401 \\
    2023-27478 & libmemcached-awesome & CWE-200 & 2022-47938 & Linux Kernel & CWE-125 \\
    2022-28388 & Linux Kernel & CWE-415 & 2023-23002 & Linux Kernel & CWE-476 \\
    2023-22996 & Linux Kernel & CWE-772 & 2022-42895 & Linux Kernel & CWE-824 \\
    2021-3743 & Linux Kernel & CWE-125 & 2022-34495 & Linux Kernel & CWE-415 \\
    2022-24958 & Linux Kernel & CWE-763 & 2022-47520 & Linux Kernel & CWE-125 \\ 
    2022-30594 & Linux Kernel & CWE-863 && \\
    \hline
   
    \end{tabular}}
    \vspace{-6pt}
  \label{tab:cve}
\end{table}

\subsection{RQ6: Real-World Vulnerability Discovery}
We scraped the latest 71 vulnerabilities covering 17 CWEs from 6 critical software projects (e.g., Linux kernel) reported between 2021-2023, from the CVE/NVD database~\cite{nvd}. We use LineVul as it is the latest available vulnerability detector. \modify{Table~\ref{tab:cve}} shows the latest vulnerabilities detected by the LineVul model improved by {\tech}'s generated samples but cannot be detected by the original one. The improved LineVul found \emph{13 more CVEs}, indicating its enhanced potential for discovering real-world zero-day vulnerabilities.



%% file: discussion.tex
\section{Discussion}
\noindent
In this section, we discuss why {\tech} has good performance on generating vulnerability data and why the generated samples by {\tech} effectively improve the performance of downstream tasks.

\subsection{Vulnerability Generation Performance}
\noindent
As discussed in $\S$\ref{sec:approach}, the key designs of {\tech} are \emph{semantics-aware contextualization} and \emph{human-knowledge-enhanced edit patterns formation}. 
We dissect {\tech}'s performance merits from these aspects.

\begin{figure}[tp]
	\vspace{-0pt}
	\includegraphics[width=0.96\linewidth]{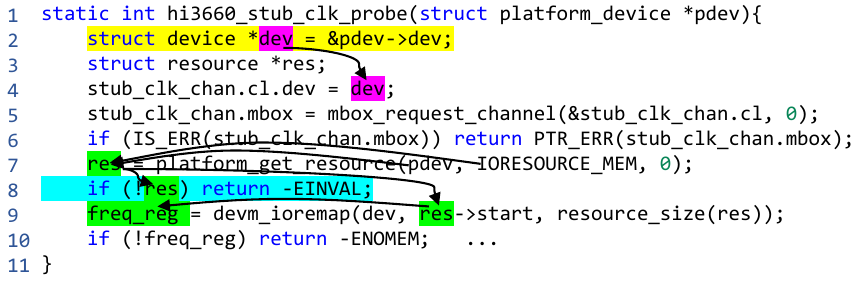}
    \vspace{-10pt}
	\caption{An example where {\tech} correctly predicts the statement at Line 8 (marked as cyan), but without value-flow-based position encoding it would incorrectly locates Line 2 (marked as yellow).}
	\label{fig:case-flow}
	\vspace{-0pt}
\end{figure}

\textbf{The value-flow-based position encoding makes the contextualization semantics-aware.} Figure~\ref{fig:case-flow} shows an example where {\tech} correctly predicts 
the vulnerability injection location due to its value-flow-based position encoding. 
The reason is that 
this encoding explicitly enhances the attention between variables that have value flow relationships. For example, in the case shown in the figure, the variable {\tt res} has value flows from {\tt pdev}, {\tt IORESOURCE\_MEM}. These variables are highly related to vulnerability injection. Thus, while they are not in the statement to be located, {\tech} pays more attention to these variables when locating the correct statement. However, the one without value-flow-based position encoding simply locates Line 2 because it has the tokens {\tt dev, pdev}, and {\tt device}, which appeared on vulnerable statements of some training samples, but that statement has nothing to do with injecting a vulnerability as per the code semantics here. {\tech} takes the advantage of the value-flow-based position encoding which makes the model semantics-aware hence better contextualization performance, \modify{
overcoming Limitation \textcircled{2} for vulnerability generation}.

\begin{figure}[tp]
	\vspace{-0pt}
	\includegraphics[width=0.96\linewidth]{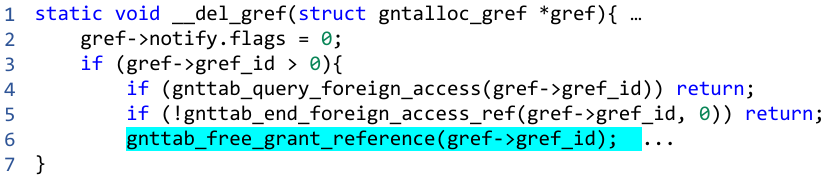}
    \vspace{-10pt}
	\caption{An example where {\tech} successfully remove the statement at Line 6 (marked as cyan) to inject a memory leak vulnerability, but without manual pattern refinement via regex it would not be able to do so.
 }
	\label{fig:case-re}
	\vspace{-0pt}
\end{figure}

\textbf{Manual pattern refinement with 
regex 
makes the patterns necessarily general hence more useful.} Figure~\ref{fig:case-re} shows an example where {\tech} successfully removes the statement at Line 6 (marked as cyan) to inject a ``memory leak (CWE-401)" vulnerability but the one without manual pattern refinement via regex
cannot. The reason is that this program uses a self-defined {\tt free} function to release the pointer. Thus, the patterns mined through anti-unification may not cover the function name of this {\tt free} function. The use of self-defined functions related to memory allocation, memory initialization, memory release, safety check, and multi-threading management are common based on our manual inspection. Thus, our manually added edit patterns with regex 
shown in Table~\ref{tab:derived-patterns} make the edit patterns necessarily more general, improving the performance of {\tech} for vulnerability generation, \modify{which effectively overcomes Limitation \textcircled{1} for vulnerability generation}. 

\begin{figure}[tp]
	\vspace{-0pt}
	\includegraphics[width=0.96\linewidth]{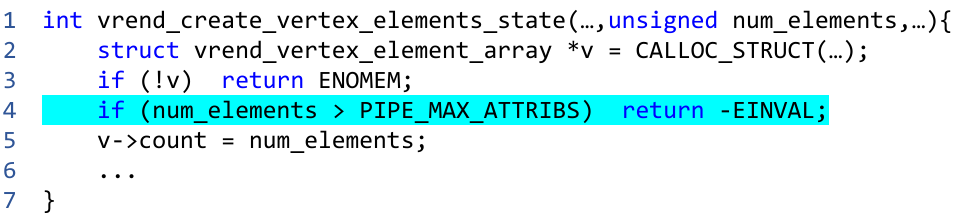}
    \vspace{-10pt}
	\caption{An example where {\tech} successfully remove the if statement at Line 4 (marked as cyan) to inject an improper input validation (CWE-20) vulnerability, but without pattern mutation it would not be able to inject correctly.}
	\label{fig:case-mutation}
	\vspace{-0pt}
\end{figure}

\textbf{Pattern diversification makes the patterns more diverse hence more applicable.}
Figure~\ref{fig:case-mutation} shows an example where {\tech} successfully 
injects a CWE-20 vulnerability due to its pattern diversification. 
The reason is that the input validation checking at Line 4 may be very different in different programs. It is difficult for the patterns extracted from 
existing examples 
to cover all of them. Yet based on our \emph{human knowledge}, the return value {\tt -EINVAL} has indicated that this is a value validation checking. Thus, we build pattern mutation rules (Table~\ref{tab:mutation} Row 4) such that once such return error values are in the {\tt if} statement, we do not need to match the condition any more. Therefore, the pattern diversification integrates our human knowledge, making the edit patterns more diverse and applicable hence improving {\tech} for vulnerability generation, \modify{
overcoming Limitation \textcircled{1} 
of peer approaches.}

\subsection{Usefulness of Generated Samples}
\modify{In $\S$\ref{sec:usefulness}, we show that {\tech}-generated samples are high-quality and thus are effective for improving various downstream vulnerability analysis tasks. Specifically, our generated samples have merits in four main aspects of high dataset quality:}

\modify{\textbf{Dataset size.} Training dataset size is crucial for training a deep learning model well. Almost all the DL-based code analysis techniques use more than 10,000 samples to train the models~\cite{bi2023benchmarking}. However, previous works~\cite{nong2022generating,nongvulgen} like VulGen can only generate a small number of vulnerability samples (e.g., <1,000) or their quality plummets. In contrast, {\tech} can generate large-scale quality vulnerability samples in a short time. The large number of generated samples allows the DL model to learn relevant knowledge in general. }

\modify{\textbf{Complexity.} Some previous vulnerability analyses use synthetic datasets to train their models~\cite{li2018vuldeepecker,li2021sysevr}. However, synthetic samples are usually very simple and make the trained models not generalizable to real-world vulnerability analysis~\cite{chakraborty2022deep}. This is also confirmed in our experiments in $\S$\ref{ssec:downstream}. In contrast, {\tech}'s generated samples are based on real-world normal samples, thus the generated samples are as complex as real-world vulnerability data and ensure the model learns the knowledge for challenging vulnerability analysis.}

\modify{\textbf{Noise.} Another important aspect of the usefulness is noise.
While VulGen~\cite{nongvulgen} can generate vulnerable samples, the success rate is low (41.33\%), making the samples unready to use. In contrast, {\tech}'s generated samples achieve a 90.13\% success rate, which can be used to improve the model training of downstream tasks directly.}

\modify{\textbf{Diversity.} Previous works~\cite{zhang2021framework,zenong281412} only generate one or a few number types of vulnerabilities, which cannot train vulnerability analysis models in general. In contrast, {\tech} is already able to generate vulnerability samples of diverse types (CWEs) while spanning a variety of (238) projects, making its augmented model training for downstream vulnerability analysis effective.}

\section{Threats to Validity}
\modify{A possible threat to internal validity is that {\tech} may have implementation errors. We mitigate it by inspecting code carefully, doing unit testing when implementing each module, and using a small dataset to test before official experiments. Another threat is that the manual inspection for the success samples may be inaccurate. We mitigate it by writing exploits for some of the samples and confirming results by multiple authors via cross validation.}

\modify{The main external validity threat is that the dataset we use may not represent real-world vulnerability data distribution. We mitigate it by using manually confirmed real-world dataset like CVE/NVD and removing any dataset overlaps during evaluation.}

%% file: related.tex
\section{Related Work}
\modify{{\bf Vulnerability dataset curation.} SARD~\cite{black2017sard} and SATE IV~\cite{okun2013report} are synthetic datasets containing 60K+ vulnerability samples, while BigVul~\cite{fan2020ac} and CVEFixes~\cite{bhandari2021cvefixes} are real-world 
datasets containing much less (<10K) samples. 
FixReverter~\cite{zenong281412} uses manually derived patterns to inject vulnerabilities to existing code. VulGen~\cite{nongvulgen} injects vulnerabilities by addressing \textit{where} and \textit{how} to inject separately. However, the label precision of these generated datasets is low. In contrast, {\tech} is the most accurate with >90\% precision.} \revise{Apocalypse~\cite{roy2018bug} uses formal techniques to automatically inject  bugs in large software code bases. Fuzzle~\cite{lee2022fuzzle} synthesizes buggy programs by encoding moves in a maze as chains of function calls. 
In comparison, {\tech} leverages value-flow-based deep representation learning and human knowledge for vulnerability injection. 
}

\modify{{\bf Source-code pre-training.} Deep learning model pre-training has been widely employed. BERT~\cite{devlin2018bert} is a pre-trained model for natural languages. CodeBERT~\cite{feng2020codebert}, CodeT5~\cite{wang2021codet5}, and SPT-Code~\cite{niu2022spt} borrow the idea from BERT to pre-train models for programming languages. In contrast, {\tech}'s pre-training is task-specific, explicitly geared toward our fine-tuning task for vulnerability injection. 
}

\modify{{\bf Human knowledge integration.} 
Integrating human knowledge into learning-based approaches has been explored. Liu et al.~\cite{liu2021combining} combine GNN with human knowledge to detect smart contract vulnerabilities. ComNet.~\cite{gao2018comnet} integrates human knowledge into DL to improve orthogonal receivers. DATGAN~\cite{lederrey2022datgan} integrates DL with expert knowledge to generate tabular data. In comparison, {\tech} integrates human knowledge into vulnerability injection patterns to boost vulnerability generation.}




%% file: conclusion.tex
\section{Conclusion}\label{sec:conclusion}
We presented {\tech}, a novel 
technique for large-scale generation of high-quality vulnerable program samples. {\tech} generates such samples 
using vulnerability-introducing code edit patterns. These patterns are initially extracted from real-world vulnerability fixes, augmented by manually defined additional patterns, and diversified through manually derived pattern mutation rules according to human knowledge about real-world vulnerabilities. 
{\tech}'s design also features a semantics-aware contextualization Transformer to identify right injection contexts, which is customized by value-flow-based position encoding and pre-trained against new objectives to facilitate learning syntactic and contextual structures of code.  
With this novel design, {\tech} largely outperforms 
all of the state-of-the-art peer approaches in terms of 
the quality of generated samples. 
We also contribute a large vulnerability dataset 
resulting from {\tech}'s in-the-wild 
sample production. We further demonstrated 
the practical usefulness of this dataset via the 
substantial improvement it brought to vulnerability detection, localization, and repair, 
and its ability to help find more real-world vulnerabilities (CVEs).
